\documentstyle[prl,aps,multicol,psfig]{revtex}

\begin{document}
%\twocolumn[\hsize\textwidth\columnwidth\hsize\csname @twocolumnfalse\endcsname
\title{Conversion of free magnetic polaron into vortex lattice in diluted magnetic
semiconductors in quantizing magnetic fields.}
\author{V.A.Stephanovich}
\address{Institute of Mathematics, University of Opole, Oleska 48, 45-052, Opole, Poland\\
and Institute of Semiconductor Physics NAS of Ukraine, Kiev, Ukraine}
\date{\today}
\maketitle

\begin{abstract}
We show that in strong (quantizing) magnetic fields "ordinary" free magnetic
polaron in diluted magnetic semiconductors (of type $\rm {A^{2}MeB^{6}}$, where $%
{\rm Me=Fe, Mn}$) exists in the form of vortex lattice quite similar to that in type II
superconductors (Abricosov vortex lattice). The region of external
parameters (like external magnetic field and temperature), where such
lattice exists, is determined from the condition that lattice dimension is
less or equal to polaron localization radius.
\end{abstract}

%]
\begin{multicols}{2}

\narrowtext

Diluted magnetic semiconductors (DMS) can be regarded as an
ensemble of magnetic impurities Fe, Mn, embedded in the lattice of host  $\rm {A^{2}B^{6}}$
semiconductor. The concentration of  impurities is such that direct exchange interaction
between impurities does not occur so that aforemetioned substances is in the paramagnetic
phase.  

Last decade considerable experimantal and theoretical efforts have been
spared for investigation of self-localized states of carriers in DMS 
(see, e.g.\cite{Di,Wo,rs,hwv,ses}). Such self-localized state of conduction electron (hole) 
is called  free magnetic polaron (MP) and can be regarded as a region 
(of polaron radius $\rho$) of correlated spins of magnetic ions, where the 
interaction between spins of magnetic impurities is mediated by the electron (hole).

However, existing theories considered the properties of such MPs  in magnetic
fields (such that $\rho _H=\sqrt{\frac{\hbar c}{eH}}<\rho$, $\rho $ is MP
localization radius, $H$ is external magnetic field), where diamagnetic
effects (Landau quantization) do not manifest themselves (see \cite{ses}).
It turns out, however, that shape and physical properties of MP may change
drastically  if we include aforementioned diamagnetic contribution into consideration.

In this paper we show that MP in quantizing magnetic fields exists as a lattice of vortices,
very similar to Abricosov vortex lattice in type II superconductors \cite{abr}. 
We have shown that formation of the vortex lattice is energetically favourable as compared
to the single MP. Our analysis show that general scenario of vortex lattice formation is as follows. 
When magnetic field exceeds some threshold value 
\begin{equation}
H_{c1}=\frac{\hbar c}{e \rho ^2},\label{elka}
\end{equation}  
the initial spherically symmetric "paramagnetic" MP 
(see \cite{ses} for detailed investigation of this MP properties; 
$\rho$ is a function of magnetic ions concentration, temperature and external magnetic field) 
looses its sprerical symmetry and gains cylindrical one. This "diamagnetic" MP 
is localized in a plane perpendicular to
magnetic field direction (i.e. in the plane of Landau quantization). At further increase of 
magnetic field this MP splits into several vortices, that tend to organize into vortex lattice
\cite{abr,kra}.

We can draw an "analogy" between such "diamagnetic" MP and vortex in 
type II superconductor. While magnetic field penetrates only to the vortex core 
in superconductor, giving rise to normal (nonsuperconducting) phase there,
the core region of MP consists of coherently oriented correlated 
spins of magnetic impurities, giving rise to (local) magnetically 
ordered phase. 

At very high magnetic fields $H \sim H_{c2}$ the "ferromagnetic" cores of the 
vortices merge so that magnetic impurities do not "need" the electron (hole) to
interact with each other. In this region of magnetic fields the amplitude of MP
wave function become small so that $H_{c2}$ can be determined by the usual
procedure of linearization \cite{abr} of corresponding equation.

In the present paper we give the estimations of $H_{c1}$ and $H_{c2}$ rather then
detailed (standard) calculation of vortex lattice properties.

Consider first the model of free MP. We start from the Hamiltonian of electron
(hole), coupled to subsystem of magnetic ions in DMS. These magnetic ions
randomly substitute $N_l$ cations at the sites ${\vec R}_j$ ($j=1,2,...N_l$)
of a cubic crystal lattice. Under the assumption that corresponding energy
band is isotropic and nondegenerate, we write the Hamiltonian of this system
in the effective mass approximation
\end{multicols}
\widetext
\noindent\rule{20.5pc}{0.1mm}\rule{0.1mm}{1.5mm}\hfill
\begin{equation}
{\cal {H}} =\frac{{\vec p}^2}{2m^{*}}+g_e\mu _B{\vec H}{\vec S}_e+
\sum_{j=1}^{N_l}\left[ -J\Omega \delta \left( {\vec r}-{\vec R}_j\right) {%
\vec S}_l^j{\vec S}_e+{\cal {H}}_l^j\left( g_l\mu _B{\vec H}\right) \right] +%
{\cal {H}}_{ll},  \label{eq1} 
\end{equation}
\hfill\rule[-1.5mm]{0.1mm}{1.5mm}\rule{20.5pc}{0.1mm}
\begin{multicols}{2}
\narrowtext
\noindent
where ${\vec p},m^{*},g_e,{\vec S}_e$ and ${\vec r}$ are momentum operator,
effective mass, band g-factor, spin operator, and coordinate of electron (hole); $%
g_l$ and ${\vec S}_l^j$ are g-factor and spin operator of the magnetic ion
localized at lattice site ${\vec R}_j$, $J$ is carrier-ion exchange
interaction constant, $\Omega $ is unit cell volume and $\mu _B$ is Bohr
magneton. The first two terms in (\ref{eq1}) determine the kinetic and
Zeeman energy of an electron. The next two terms determine the carrier-ion
exchange interaction and the sum of magnetic ions Hamiltonians in the field $%
{\vec H}$ (see \cite{ses} for details).

Choosing electron wave function as a product of coordinate and spin parts
and averaging over random impurities positions in a mean field approximation
(see \cite{ses} for details), we obtain following energy functional for a
free magnetic polaron.
\end{multicols}
\widetext
\noindent\rule{20.5pc}{0.1mm}\rule{0.1mm}{1.5mm}\hfill
\begin{eqnarray}
E_\sigma  &=&\int \Biggl\{\frac 1{2m_z}\psi ^*\left( %
\hat {p}_z-\frac ecA_z \right) ^2 \psi + 
\frac 1{2m_{\perp }}\psi ^*\left(\hat {\vec p}_{\perp} -\frac ec{%
\vec A}_{\perp } \right) ^2 \psi \Biggr\}d^3r  
+g_e\mu _BH\sigma + \label{eq2}  \\ 
&+&n_l\int \left[ E_1\left( g_l\mu _BH+Q\sigma \right) -E_1\left( g_l\mu
_BH\right) \right] d^3r,  \nonumber
\end{eqnarray}
\hfill\rule[-1.5mm]{0.1mm}{1.5mm}\rule{20.5pc}{0.1mm}
\begin{multicols}{2}
\narrowtext
\noindent
where $\sigma =\pm \frac 12$ are the eigenvalues of the operator projecting
the electron spin onto ${\vec H}$ and

\begin{equation}
Q\left( {\vec r}\right) =-J\Omega \left| \psi \left( {\vec r}\right) \right|
^2  \label{eq3}
\end{equation}
is the effective exchange field, $E_1$ depends on the specific type of SMS
magnetic ions paramagnetism (see below), 
$ \hat {p}_z \equiv -i\hbar \frac \partial {\partial z}$, 
$\hat {\vec p}_{\perp}\equiv -i\hbar ({\vec e}_x\frac \partial {%
\partial x}+{\vec e}_y\frac \partial {\partial y}),\ {\vec A}_{\perp }=
{\vec e}_xA_x+{\vec e}_yA_y$. 

For SMS of type $\rm {A_{1-x}^2Fe_xB^6}$ with Van Vleck paramagnetism $E_1$ has
the form \cite{sem}
\begin{equation}
E_{1VV}(x)=\Delta \varepsilon -\left[ (\Delta \varepsilon )^2+4x^2\right]
^{1/2},  \label{eq4}
\end{equation}
where $\Delta \varepsilon $ is the splitting of the spin-orbit multiplet of
the magnetic ion (see \cite{sem} for details). For SMS of type $%
\rm {A_{1-x}^2Mn_xB^6}$ with orientational paramagnetism we have for $E_1$%
\begin{equation}
E_{1OR}\left( x\right) =\frac 1\beta \log \frac{\sinh \left[ \left( S+\frac 1%
2\right) \beta x\right] }{\sinh \left( \frac{\beta x}2\right) },  \label{eq5}
\end{equation}
where S is magnetic ion spin, $\beta =\left( k_BT\right) ^{-1}.$ 

It is seen that  (\ref{eq2}) looks like  Ginzburg-Landau (GL) functional for
superconductors. The difference is more complicated nonlinearity. It is seen
that expansion of Eqs (\ref{eq4}) and (\ref{eq5}) in powers of $\left|
\psi \left( {\vec r}\right) \right| ^2$ up to $\left| \psi \left( {\vec r}%
\right) \right| ^4$ put formal one-to-one correcpondence with GL functional.

Variation of (\ref{eq2}) with respect to $\psi ^*$ gives following Schr\"{o}dinger
equation for polaron wave function
\begin{equation}
\frac 1{2m_{\perp }}\left(\hat {\vec p}_{\perp} -\frac ec{%
\vec A}_{\perp } \right) ^2 \psi+n_l \frac {\partial E_1}{\partial \left| \psi \right|^2 }\psi =0.
\label{eq6}
\end{equation}
Here we omit the unimportant dependence on $z$, since the electron motion is not 
quantized in this direction. Boundary conditions for MP are similar to those for
GL superconducting order parameter and require that $\psi =0$ at infinity.

It is convenient to show the existense of vortex lattice solutions of 
corresponding differential equations from the side of $H \sim H_{c2}$,
where $\left|\psi \left( {\vec r}\right) \right| ^2$ is small. For that we
linearize  (\ref{eq6}) (with respect to  (\ref{eq4}) and  (\ref{eq5})) 
in $\psi $. This gives
\begin{equation}
\frac 1{2m_{\perp }}\left( -i\hbar \frac {\partial}{\partial y}+\frac ec Hx\right)^2 \psi+
\lambda \psi =0, \label{eq7}
\end{equation}
where we choose following gauge of vector potential
\begin{equation}
A_y=Hx,\ H\equiv H_z,  \label{eq3a}
\end{equation}
\begin{eqnarray}
\lambda _{VV}&=&J \Omega n_l \sigma \frac{4g_l\mu _BH}{\sqrt{(\Delta \varepsilon)^2+%
4(g_l\mu _BH)^2}},\nonumber \\
\lambda _{OR}&=&-J \Omega n_l\sigma SB_S(S\beta g_l\mu _BH), \label{eq8}
\end{eqnarray}
$B_S(x)$ is a Brillouin function for spin S:
\[
SB_S(Sx)=\left(S+\frac 12\right)\coth \left( S+\frac 12 \right)x-\frac 12 \coth \frac x2.
\]
It is well-known (see, e.g. \cite{kvmex})
that localized solutions of (\ref{eq7}) exist if $\lambda$ equals 
to one of eigenvalues 
\[
\lambda = -(n+\frac 12)\hbar \frac{eH}{m_{\perp}c}.
\] 

Lowest eigenvalue determines $H_{c2}$ so that we obtain following equations for $H_{c2}$
for the case of Van Vleck
\begin{mathletters}
\begin{equation}
J x_l \sigma \frac{4g_l\mu _BH}{\sqrt{(\Delta \varepsilon)^2+%
4(g_l\mu _BH)^2}}=-\frac 12 \hbar \frac{eH}{m_{\perp}c} \label{eq9a}
\end{equation}
and orientational paramagnetism
\begin{eqnarray}
&&\sigma J x_lSB_S(S\beta g_l\mu _B{H_{c2OR}})-\frac{\mu _B{H_{c2OR}}}{\mu } = 0,\nonumber \\ 
&&\mu =\frac {m_{\perp}}{m_0},\ x_l=n_l \Omega, \label{eq9b}
\end{eqnarray}
\end{mathletters}
where $m_0$ is a free electron mass.

To have feeling of the numerical values of $H_{c2}$ for some specific DMS, we use two examples.
First one is $\rm{Zn_{1-x}Fe_xSe}$ with Van Vleck paramagnetism of Fe. Its parameters are \cite{vv}:
$m^{*}_e=0.14m_0$; 
$m^{*}_h=1.2m_0$; $J_e=0.22 eV$; $J_h=-1.6 eV$; $\Delta \varepsilon =1.8 meV$;
unit cell volume $\Omega=45.6 \AA ^3$. It is sufficient for our estimations to 
put $m^{*}_{e,h}=m_{\perp e,h}$. 

Second one is $\rm{Cd_{1-x}Mn_xTe}$ with orientational paramagnetism of Mn. Parameters of this 
solid solution are following \cite{ori}: $m^{*}_e=0.096m_0$; 
$m^{*}_h=0.48m_0$; $J_e=0.22 eV$; $J_h=-0.88 eV$; lattice constant $a=6.48\AA$; 
unit cell volume $\Omega=68.06 \AA ^3$. 

It is seen from (\ref{eq9a}) that this equation has solution only if $J\sigma <0$,
i.e. for vortex lattice to occur for electron MP ($J_e>0$), we should have $\sigma =-1/2$;
for hole MP $\sigma = 1/2$. We have from (\ref{eq9a})
\begin{eqnarray}
&&\frac 1{\sqrt{1+4h_{c2VV}^2}}=\frac 14 \frac {\Delta \varepsilon}{\mu g_lx_l|J\sigma|},\nonumber \\
&&h=\frac {g_l\mu _BH}{\Delta \varepsilon}. 
 \label{eq9c}
\end{eqnarray}
Since $\Delta \varepsilon /J <<1$, the solution of (\ref{eq9c}) is at $h >>1$. In this case
we have from (\ref{eq9c})
\begin{equation}
H_{c2VV}  =(1.7\cdot 10^4 \ {\rm {Tesla}})\ \mu x_l |J_0|,\ J_0=J/1eV \label{eq9d} 
\end{equation}
It was shown earlier \cite{ses} that in $\rm{Zn_{1-x}Fe_xSe}$ electron cannot autolocalize.
So, we give estimate of $H_{c2}$ for hole only.
From (\ref{eq9d}) (at typical value $x_l=4\%$) we obtain, that for 
hole $H_{c2VV}^h\approx 1300$ Tesla. 

The solution of Eq.\ (\ref{eq9b}) exists at $J\sigma >0$. In dimensionless variables
\[
\frac {g_l\mu _BH}{S\mu J x_l}=h,\ \frac {3k_BT}{JS(S+1)\mu x_l}=\tau
\] 
equation (\ref{eq9b}) can be rewritten as
\[
h_{c2OR}=B_S\left(\frac {3h_{c2OR}}{\tau (S+1)}\right)
\]
Its asymptotic values are as follows
\begin{eqnarray}
&&k_BT_c=\frac J3 S(S+1)\mu x_l,\nonumber \\ 
&&g_l\mu _BH_{c2OR}(T=0)=JS\mu x_l. \label{eq9e}
\end{eqnarray}
The "universal" (i.e. independent of $J$ and $x_l$) temperature dependence of 
$H_{c2OR}$ is reported in Fig.1. Our estimations show that elctron cannot 
autolocalize also in orientational DMS $\rm{Cd_{1-x}Mn_xTe}$.
Equation (\ref{eq9e}) permits to estimate $H_{c2OR}^h$ from above. 
Taking into account that MP in orientational DMS can be formed
at $x_l>17\%$, we obtain (for $S=5/2$ and $x_l=20\%$) 
for hole $H_{c2OR}^h\approx 1800$ Tesla.

\begin{figure}[th]
\vspace*{-5mm}
\centerline{\centerline{\psfig{figure=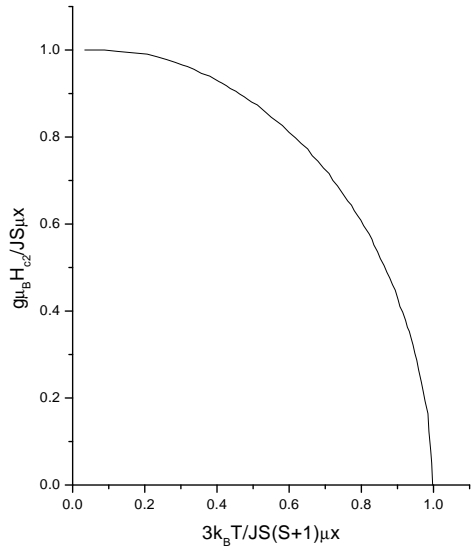,width=0.8\columnwidth}}}
%\vspace*{0.5mm}
\caption{"Universal" temperature dependence of $H_{c2}$ for orientational DMS.}
%\label{diagram}
\end{figure}

The situation with $H_{c1}$ is quite similar to that in type II superconductors. In this case
MP plays a role of single vortex. Our analysis show that at $H<H_{c1}$ MP exists in
3D spherically symmetrical form while at $H>H_{c1}$ it has 2D character being localized in the
plane perpendicular to the direction of the external magnetic field. So, for $H_{c1}$ 
estimation it is sufficient to consider spherically symmetrical MP, substituting its radius
of localization to Eq.\ (\ref{elka}). Since at $H \sim H_{c1}$
we should keep nonlinearity, we cannot solve Eq.\ (\ref{eq6}) analytically. 
However, MP localization radius $\rho$ can be
quite accurately calculated variationally with the trial MP wave function in the 
simplest possible form \cite{ses}
\begin{equation}
\psi = \left(\frac {2}{\pi \rho ^2}\right)^{\frac 34} \exp%
{\left(-\frac{r^2}{\rho ^2}\right)}.\label{eq10} 
\end{equation} 

In Van Vleck DMS such variational calculation had been done in \cite{ses}. For $H_{c1}$
we obtain from (\ref{elka})
\begin{equation}
H_{c1}=\frac {87.4 \ {\rm Tesla}}{\rho _0^2},\ \rho _0=\rho \sqrt{\frac{\pi }{2}}%
\left(\frac{\Delta \varepsilon}{J \Omega}\right)^{\frac 13} \label{eq11}
\end{equation}
For $x_l=4\%$ variational calculation \cite{ses} gives $\rho _0=0.75$ and 
$H_{c1VV}^h = 155$ Tesla. 

Thus at $x_l=4\%$  vortex lattice for hole MP in Van Vleck
DMS possibly occurs in wide range of magnetic fields 
$155 \ {\rm Tesla} <H< 1300 \ {\rm Tesla}$.

We have for $H_{c1}$ in the case of orientational DMS
\begin{equation}
H_{c1}=\frac {6000 \ {\rm Tesla}}{\rho _0^2},\ \rho _0=\rho \sqrt{\frac{\pi }{2}}%
\left(\frac{1}{\Omega}\right)^{\frac 13} \label{eq12}
\end{equation}
For $x_l=20\%$ and $T=10K$ $\rho _0=14$ and $H_{c1OR}^h = 30$ Tesla. 

Thus at $x_l=20\%$  and low temperatures vortex lattice for hole MP in orientational
DMS possibly occurs for magnetic fields 
$30 \ {\rm Tesla} <H< 1800 \ {\rm Tesla}$. This interval is
by order of magnitude the same as that in Van Vleck DMS. The difference is that in
orientational DMS we may "adjust" (e.g. getting lower) critical fields
by temperature. 

We have shown that quantizing magnetic field "forces" free spherically
symmetric MP in DMS to split into the vortex lattice very similar to that
in type II superconductors. Vortex lattice possibly exists at magnetic fields
much higher than those for type II superconductors.
It occurs for $J\sigma <0$ for 
Van Vleck DMS and for $J\sigma >0$ for orientational ones. The values of 
$H_{c1}$ and $H_{c2}$ for orientational DMS can be "adjusted"
by temperature, which is impossible for the Van Vleck DMS. This permits to hope,
that vortex lattice can be detected more easily in DMS with orientational paramagnetism.
As to Van Vleck DMS,in spite of the fact that $H_{c1}$ and $H_{c2}$ are high,
%Although this phenomenon occurs at very high magnetic fields , 
they are accessible by modern experimental equipment. We think also, that it is possible
to find Van Vleck DMS, where considered phenomena can occur at lower magnetic fields.

\end{multicols}


\begin{references}
\bibitem{Di}   T. Dietl and J. Spalek, \prb {\bf 28}, 1548 (1983). 

\bibitem{Wo}  P.A. Wolff, Semicond. Semimet. 25, 413, Academic, New
York (1988). 

\bibitem{rs}   S.M. Ryabchenko and Yu. G. Semenov, Zh. Eksp. Teor. Fiz.
{\bf 84}, 1419 (1983) [Sov. Phys. JETP {\bf 57} , 825 (1983)]; Fiz. Tv. Tela (Leningrad)
{\bf 26}, 3347 (1984) [Sov. Phys. Solid State {\bf 26}, 2011 (1984)]. 

\bibitem{hwv}  D. Heiman, P.A. Wolff, and J. Warnok, \prb {\bf 27}, 4848,
(1983). 

\bibitem{ses}  Yu.G. Semenov, V.A. Stephanovich, Zh. Eksp. Teor. Fiz.
{\bf 101}, 1024 (1992) [Sov. Phys. JETP {\bf 74} , 549 (1992)]. 

\bibitem{abr} A.A.Abrikosov Zh. Eksp. Teor. Fiz, {\bf  32}, 1442 (1957)
[Sov. Phys. JETP {\bf 5} , 1174 (1957)].

\bibitem{kra} W.H.Kleiner, L.M.Roth, S.H.Autler, Phys. Rev. {\bf 133}, A1266
(1964).

\bibitem{sem}  Yu.G. Semenov, Fiz. Tekh. Poluprovodn. {\bf 21}, 1802 (1987)
[Sov. Phys. Semicond. {\bf 21}, 1092 (1987)].

\bibitem{kvmex} L.D.Landau, E.M.Lifshits Quantum Mechanics, Pergamon Press,1977.

\bibitem{vv} A.Twardowski, P.Glod, W.J.M. de Jonge, M.Demianiuk Sol. State Comm.,
{\bf 64}, 63 (1987), P.Lawaetz \prb {\bf 4}, 3460 (1971). 

\bibitem{ori} J.A.Gaj, W.Grieshaber, C.Boden-Deshayes, J.Cibert, G.Feuillert, 
Y. Merle d'Aubigne and A.Wasiela. \prb {\bf 50}, 5512 (1994).
\end{references}
\end{document}